\documentclass[11pt]{article}
\usepackage{amsfonts}
\usepackage{lscape}
\usepackage{latexsym,amsmath,amsthm}
\usepackage{amssymb,array}
\graphicspath{{figures/}}
\usepackage{subfigure}
\usepackage{fancyhdr}
\usepackage{multicol}
\usepackage{float}
\usepackage{csquotes}
\makeatletter 
\RequirePackage[dvips]{graphicx} 
\newtheorem{t1}{Theorem}[section]
\newtheorem{l1}{Lemma}[section]

\begin{document}
\title{\textbf{Study on estimators of the PDF and CDF of the one parameter polynomial exponential distribution}}
\author{Indrani Mukherjee$^1$, Sudhansu S. Maiti$^1$\footnote{Corresponding author e-mail:
dssm1@rediffmail.com} and Vijay Vir Singh$^2$ \\
$^1$Department of Statistics, Visva-Bharati
University, Santiniketan-731 235, West Bengal, India\\
$^2$Department of Mathematics, Faculty of science, Yusuf Maitama Sule University Kano, Nigeria}
\date{}
\maketitle
\begin{center}
Abstract
\end{center}
In this article, we have considered one parameter polynomial exponential distribution. The exponential, Lindley, length-biased Lindley and Sujatha distribution are particular cases. Two estimators viz, MLE and UMVUE of the PDF and the CDF of the OPPE distribution have been discussed. The estimation issue of the length-biased Lindley and Sujatha distribution have been considered in detail. The estimators have been compared in MSE sense. Monte Carlo simulations and real data analysis are performed to compare the performances
of the proposed methods of estimation.
\vspace{0.5cm}


\newpage
\section{Introduction}
\ Statistical models are very useful in describing and predicting real-world phenomena. Recent developments focus on defining new families that extend well-known distribution and at the same time providing greater flexibility in modelling data in practice. Many well known lifetime distributions for modelling lifetime data such as exponential, Gamma, Weibull, etc. have been extensively studied.

\par Let $X$ is a random variable taking values $(0,\infty)$. So the distribution of $X$ may be absolutely continuous or discrete. The probability density function (PDF) and the cumulative distribution function (CDF) of the Lindley distribution [see, Lindley (1958)] are given by
\begin{equation}\label{ch4lindpdf}
f(x)=\frac{\theta^2}{1+\theta}(1+x)e^{-\theta x}, ~ x,~ \theta >0
\end{equation}
and 
\begin{equation}\label{ch4lindcdf}
F(x)=1-\frac{(1+\theta +\theta x)}{1+\theta} e^{-\theta x},~ x,~ \theta >0,
\end{equation}
respectively.

\vspace{0.1in}
The exponential distribution is closed in form to the Lindley distribution given in $(\ref{ch4lindpdf})$. 
The PDF and the CDF of the exponential distribution are given by
\begin{eqnarray}\label{ch4exppdf}
f(x)&=&\theta e^{-\theta x},~ x,~ \theta >0
\end{eqnarray}
and 
\begin{eqnarray}\label{ch4expcdf}
F(x)&=&1-e^{-\theta x},~ x,~ \theta >0,~\mbox{respectively}.
\end{eqnarray}
\par Many of the mathematical properties (e.g., the mode of the distribution, moments, skewness and kurtosis measures, cumulants, failure rate and mean residual life, mean deviation, entropies, etc.) are more flexible than those of the exponential distribution. For the exponential distribution, some of the properties are constant, usually not appropriate assumptions in reality whereas for the Lindley distribution there is scope to vary [see, Ghitany et al. (2008)].

\par The Lindley distribution is one way to describe the lifetime of a process or device. It can be used in a wide variety of fields, including biology, engineering and medicine. Ghitany et al. (2008) fitted this distribution to the waiting times for getting service of bank customers data. Ghitany et al. (2011) stated that it is especially useful for modeling in mortality studies. Mazucheli and Achcar (2011) discussed the applications of the Lindley distribution to competing risk lifetime data. Mukherjee and Maiti (2014) has used this distribution for constructing an acceptance sampling plan for the variable. Maiti et al. (2014) applied it in the context of describing a new process capability index.

\par It has been generalized by a host of authors. To mention a few, Zakerzadeh and Dolati
(2009), Bakouch et al. (2012), Shanker and Ghebretsadik (2013), Elbatal et al. (2013), Ghitany
et al. (2013) among others. Bouchahed and Zeghdoudi (2018) has proposed a new and unified
approach in generalizing the Lindley's distribution. They investigated some structural properties like moments, skewness, kurtosis, median, mean deviations, Lorenz curve, entropies and
limiting distribution of extreme order statistics; reliability properties like reliability function,
hazard rate, stress-strength reliability, stochastic ordering; and estimation methods like the
method of moment and maximum likelihood. We call the proposed distribution of Bouchahed
and Zeghdoudi (2018) as the one parameter polynomial exponential (OPPE) distribution. 

\par The PDF of a random variable $X$ of the OPPE distribution can be written as
\begin{equation}
f(x)=h(\theta)p(x)e^{-\theta x}, ~ x,~ \theta >0,
\end{equation}
where, $h(\theta)=\frac{1}{\sum_{k=0}^{r}a_{k}\frac{\Gamma(k+1)}{\theta^{k+1}}}$ ,
$p(x)={\sum_{k=0}^{r}a_{k}{x^k}}.$

\par The distribution can also be written as
\begin{eqnarray}\label{ch4oppepdf}
f(x)&=&h(\theta){\sum_{k=0}^{r}a_{k}x^{k}e^{-\theta x}}\nonumber\\
&=&\frac{{\sum_{k=0}^{r}a_{k}\frac{\Gamma(k+1)}{\theta^{k+1}}f_{GA}(x;k+1,\theta)}}{\sum_{k=0}^{r}a_{k}\frac{\Gamma(k+1)}{\theta^{k+1}}},
\end{eqnarray}
where $f_{GA}(x;k+1,\theta)$ is the PDF of a gamma distribution with shape parameter $(k+1)$ and scale parameter $\theta,$ and $a_k$'s are non-negative constants. The distribution is a finite mixture of $(r+1)$ gamma distributions.

\par The CDF is given by
\begin{equation}\label{ch4oppecdf}
F(x)=1-\left(\frac{{\sum_{k=0}^{r}\frac{a_{k}\Gamma(k+1) \Gamma{(k+1,\theta x)}}{\theta^{k+1}}}}{{\sum_{k=0}^{r}a_{k}\frac{k!}{\theta^{k+1}}}}\right),~ x,~ \theta >0,
\end{equation}
 where $\Gamma(m,x)=\frac{1}{\Gamma(m)}\int_{x}^{\infty}e^{-u}u^{m-1}du$.

\vspace{0.2in}
Some special cases are as follows:
\begin{enumerate}
\item[(a)] $r=0,~a_0=1$ gives the exponential distribution,
\item[(b)] $r=1,~a_0=1,~a_1=1$ gives the Lindley distribution,
\item[(c)] $r=2,~a_0=1,~a_1=0,~a_2=1$ gives the Akash distribution [c.f. Shanker (2015)],
\item[(d)] $r=2,~a_0=1,~a_1=2,~a_2=1$ gives the Aradhana distribution [c.f. Shanker (2016b)],
\item[(e)] $r=2,~a_0=1,~a_1=1,~a_2=1$ gives the Sujatha distribution [c.f. Shanker (2016e)],
\item[(f)] $r=2,~a_0=0,~a_1=1,~a_2=1$ gives the length-biased Lindley distribution [c.f. Ayesha (2017)],
\item[(g)] $r=3,~a_0=1,~a_1=1,~a_2=1,~a_3=1$ gives the Amarendra distribution [c.f. Shanker
(2016a)],
\item[(h)] $r=4,~a_0=1,~a_1=1,~a_2=1,~a_3=1,~a_4=1$ gives the Devya distribution [c.f. Shanker
(2016c)],
\item[(i)] $r=5,~a_0=1,~a_1=1,~a_2=1,~a_3=1,~a_4=1,~a_5=1$ gives the Shambhu distribution [c.f. Shanker
(2016d)].
\end{enumerate}

\par Statisticians are most of the times interested about inferring the parameter(s) involved in the distribution. Maximum likelihood estimator (MLE) and Bayes estimate of the parameter has been focused by the authors. Hardly any unbiased estimator of the parameter has been studied so far and finding out uniformly minimum variance unbiased estimator (UMVUE) of the parameter seems to be intractable and consequently the comparison with any unbiased class of estimators is not being made. However, instead of studying the estimators of the parameter(s), we have scope to find out unbiased estimator of the PDF and the CDF as well as biased estimator of the same and comparison between the estimators could be made. That is why we have shifted our focus from estimation of parameter(s) to estimation of the PDF and the CDF. 

\par The estimators for the PDF, CDF or both can be used to estimate various functions, like differential entropy, R\'{e}nyi entropy, Kullback-Leibler divergence, Fisher information, reliability function, cumulative residual entropy, the quantile function, Bonferroni curve, Lorenz curve, probability weighted moments, hazard rate function, mean deviation about mean etc.

\par There are few works available relating to estimation of the PDF and the CDF of different probability distributions. References include; Asrabadi (1990), Dixit and Jabbari (2010) and
Dixit and Jabbari (2011) - Pareto distribution; Alizadeh et al. (2015), Mukherjee et al. (2016)
- generalized exponential distribution; Bagheri et al. (2014) - generalized exponential-Poisson
distribution; Bagheri et al. (2016b) - Weibull extension model; Bagheri et al. (2016a) - exponentiated Gumbell distribution; Jabbari and Jabbari (2010) - exponentiated Pareto distribution;
Maiti and Mukherjee (2018) - Lindley distribution; Tripathi et al. (2017b) - Generalized Logistic distribution; Tripathi et al. (2017a) - exponentiated moment exponential distribution;
Mukherjee and Maiti (2019) - lognormal distribution.

\par Organization of the article is as follows. Section $\ref{ch4mlepdfcdf}$ deals with MLE of the PDF and the CDF of the OPPE distribution. Section $\ref{ch4umvueepdfcdf}$ is devoted to finding out the UMVUE of the PDF and the CDF and their MSEs (in this case the variances). Particular case like length-biased Lindley and Sujatha distribution have been discussed in section $\ref{ch4secparticular}$. In section $\ref{ch4sim}$, simulation study results are reported and comparisons are made. Real-life data sets have been analyzed in section $\ref{ch4data}$. In section $\ref{ch4conclusion}$, concluding remarks are made based on the findings of this article.

\section{MLE of the PDF and the CDF}\label{ch4mlepdfcdf}
\ Let $X_1, X_2,...,X_n$ be random sample of size $n$ drawn from the PDF in $\eqref{ch4oppepdf}$. Here we try to find the MLE of $\theta$ which is denoted as $\widetilde{\theta}$. The log-likelihood of $\theta$ is given by
\begin{eqnarray*}
l(\theta)&=&\ln L(\theta|X)\\
&=&n \ln h(\theta)+\sum_{i=1}^{n}\ln p(X_{i})-\theta \sum_{i=1}^{n} X_{i}
\end{eqnarray*}
Now,
\begin{eqnarray}\label{ch4mle}
\frac{dl(\theta)}{d\theta}&=&0\nonumber\\
\mbox{i.e.}~ n \frac{d}{d\theta}\left(\ln h(\theta)\right)-\sum_{i=1}^{n} X_{i}&=&0\nonumber\\
\mbox{i.e.}~ \frac{\sum_{k=0}^{r} a_{k}\frac{\Gamma(k+2)}{\theta^{k+2}}}{\sum_{k=0}^{r} a_{k}\frac{\Gamma(k+1)}{\theta^{k+1}}}-\overline{X}&=&0
\end{eqnarray}
Since, the MLE of $\theta$ is not of a closed form expression, we have to solve ($\ref{ch4mle}$) numerically to obtain the MLE of $\theta$. 
Theoretical expressions for the MSE of the MLEs are not available. MSE will be studied through simulation.

\section{UMVUE of the PDF and the CDF}\label{ch4umvueepdfcdf}
\ In this section, we obtain the UMVUE of the PDF and the CDF of the OPPE distribution. Also, we obtain the MSEs of these estimators. 
\par To derive the UMVUE of the PDF and the CDF (Theorem $\ref{ch4theoremumvuepdfcdf}$), we will use Theorem $\ref{ch4ft}$ and Lemma $\ref{ch4lemma}$.

 
\begin{t1}\label{ch4ft}
Let $X_1, X_2,..., X_n \sim f_{OPPE}(x,\theta)$. Then the distribution of $T=X_1+X_2+....+X_n$ is
\begin{eqnarray*}
f(t)&=&h^{n}(\theta)\sum_{q_0} \sum_{q_1} \ldots \sum_{q_r} c(n,q_0,q_1,\ldots ,q_r)\exp(-\theta t)\\
&&\times t^{\sum_{k=0}^r (k+1)q_k-1},~t>0
\end{eqnarray*}

 with $q_0+q_1+...+q_r=n$ and
 $c(n,q_0,q_1,\ldots ,q_r)=\frac{n!}{q_0!q_1!\ldots q_r!}\frac{\prod_{k=0}^{r}[a_k\Gamma(k+1)]^{q_k}}{\Gamma(\sum_{k=0}^r (k+1)q_k)}$.
\end{t1}

\begin{proof}
The mgf of $T$ is
\begin{eqnarray*}
M_{T}(t)&=&h^{n}(\theta) \left[{\sum_{k=0}^{r} a_k \frac{\Gamma({k+1})}{\theta^{k+1}} (1-\frac{t}{\theta})^{-(k+1)}} \right]^{n}\\
&=&h^{n}(\theta)\left[ a_{0}\frac{\Gamma(1)}{\theta}\left(1-\frac{t}{\theta}\right)^{-1}+....+a_{r}\frac{\Gamma(r+1)}{\theta^{r+1}}\right.\\
&&\left.\times \left(1-\frac{t}{{\theta}}\right)^{-(r+1)} \right]^{n}\\
&=&h^{n}(\theta)\sum_{q_0} \sum_{q_1}\ldots \sum_{q_r}  \frac{n!}{q_0!q_1!\ldots q_r!}  \prod_{k=0}^{r}\left(a_k\Gamma(k+1)\right)^{q_k}\\
&&\times \theta^{-\sum_{k=0}^r (k+1)q_k}\left(1-\frac{t}{\theta}\right)^{-\sum_{k=0}^r (k+1)q_k}.
\end{eqnarray*}
Hence, the distribution of $T$ is
\begin{eqnarray*}
f(t)&=&h^{n}(\theta) \sum_{q_0} \sum_{q_1}\ldots \sum_{q_r}  \frac{n!}{q_0!q_1!\ldots q_r!}   \prod_{k=0}^{r} \left( a_k\Gamma(k+1)\right)^{q_k}\\
&&\times \theta^{-\sum_{k=0}^r (k+1)q_k} f_{GA}(t,\sum_{k=0}^r (k+1)q_k,\theta)\\
&=&h^{n}(\theta)\sum_{q_0} \sum_{q_1}\ldots \sum_{q_r} c(n,q_0,q_1....,q_r)t^{\sum_{k=0}^r (k+1)q_k-1} \exp(-\theta t),
\end{eqnarray*}
 where $c(n,q_0,q_1,\ldots ,q_r)=\frac{n!}{q_0!q_1!\ldots q_r!}\frac{\prod_{k=0}^{r}[a_k\Gamma(k+1)]^{q_k}}{\Gamma\left(\sum_{k=0}^r (k+1)q_k\right)}$.
\end{proof}

\begin{l1}\label{ch4lemma}
The conditional distribution of $X_{1}$ given $X_{1}+X_{2}+....+X_{n}=T$ is
\begin{eqnarray*}
f_{X_1|T}(x|t)&=&\frac{p(x)}{A_{n}(t)}\sum_{y_0} \sum_{y_1}...\sum_{y_r} c(n-1,y_0,y_1,\ldots ,y_r)\\
&&\times(t-x)^{\sum_{k=0}^r (k+1)y_k-1},~0<x<t,
\end{eqnarray*}

where
$$A_{n}(t)=\sum_{q_0} \sum_{q_1}\ldots \sum_{q_r}c(n,~q_0,~q_1,~\ldots,~q_r)t^{\sum_{k=0}^r (k+1)q_k-1}$$
and
 $$c(n-1,y_0,y_1....,y_r)=\frac{(n-1)!}{y_0!y_1!....y_r!}\prod_{k=0}^{r}\left(a_k\Gamma(k+1)\right) ^{y_{k}}\frac{1}{\Gamma(\sum_{k=0}^r(k+1)y_k)},$$with $y_0+y_1+y_2+....+y_r=n-1.$
\end{l1}
\begin{proof}
\begin{eqnarray*}
f_{X_1|T}(x|t)&=& \frac{f_{X_1}(x)f(t-x)}{f(t)}\\
&=&\frac{p(x)}{A_{n}(t)}\sum_{y_0} \sum_{y_1}\ldots \sum_{y_r} c(n-1,y_0,y_1,\ldots ,y_r)\\
&&\times (t-x)^{\sum_{k=0}^r (k+1)y_k-1}.
\end{eqnarray*}
\end{proof}

\begin{t1}\label{ch4theoremumvuepdfcdf}
Let $T=t$ be given. Then
\begin{eqnarray}\label{ch4umvuepdf}
\widehat{f}(x)&=&\frac{p(x)}{A_{n}(t)}\sum_{y_0} \sum_{y_1}...\sum_{y_r} c(n-1,y_0,y_1,\ldots ,y_r)\nonumber\\
&&\times (t-x)^{\sum_{k=0}^r (k+1)y_k-1},~0~<~x~<~t,
\end{eqnarray}
is UMVUE for $f(x)$ and
\begin{eqnarray}\label{ch4umvuecdf}
\widehat{F}(x)&=&1-\frac{1}{A_n(t)}\sum_{y_0} \sum_{y_1}...\sum_{y_r} c(n-1,y_0,y_1,\ldots ,y_r)t^{\sum_{k=0}^r(k+1)y_k}\nonumber\\
&&\times \sum_{k=0}^r a_k t^{k}I_{x/t}\left((k+1),\sum_{k=0}^r (k+1)y_k\right)\nonumber\\
&&,~0~<~x~<~t,
\end{eqnarray}
is UMVUE for $F(x)$,
where $I_{x}(\alpha ,\beta)=\frac{1}{B(\alpha ,\beta)}\int_x^1 u^{\alpha -1}(1-u)^{\beta -1}du$ is an incomplete beta function and 
$B(\alpha ,\beta)=\frac{\Gamma \alpha\Gamma \beta}{\Gamma (\alpha +\beta)}$.
\end{t1}
\begin{proof}
From Lemma $\ref{ch4lemma}$ we get UMVUE of the PDF. 
\begin{eqnarray*}
\widehat{F}(x)&=& 1-\int_x ^t \widehat{f}(m)dm\\
&=&1-\int_x ^t \frac{p(m)}{A_{n}(t)}\sum_{y_0} \sum_{y_1}...\sum_{y_r} c(n-1,y_0,y_1,\ldots ,y_r)\\
&&\times (t-m)^{\sum_{k=0}^r (k+1)y_k-1}\\
&=&1-\frac{1}{A_n(t)}\sum_{y_0} \sum_{y_1}...\sum_{y_r} c(n-1,y_0,y_1,\ldots ,y_r)t^{\sum_{k=0}^r(k+1)y_k}\nonumber\\
&&\times \sum_{k=0}^r a_k t^{k}I_{x/t}\left((k+1),\sum_{k=0}^r (k+1)y_k\right)
\end{eqnarray*} 
\end{proof}
The MSE of $\widehat{f}(x)$ is given by
\begin{eqnarray}\label{ch4msefhatx}
MSE(\widehat{f}(x))&=&E(\widehat{f}^2(x))-f^2(x)\nonumber\\
&=&\int _x^\infty \left[\frac{p(x)}{A_{n}(t)}\sum_{y_0} \sum_{y_1}...\sum_{y_r} c(n-1,y_0,y_1,\ldots ,y_r)\right.\nonumber\\
&&\left.\times(t-x)^{\sum_{k=0}^r (k+1)y_k-1}\right]^2\nonumber\\
&&\times f(t)dt-f^2(x)
\end{eqnarray}
Using Theorem $\ref{ch4ft}$, $\eqref{ch4oppepdf}$ in $\eqref{ch4msefhatx}$, we can get the value of the MSE of UMVUE of the PDF.
And the MSE of $\widehat{F}(x)$ is given by
\begin{eqnarray}\label{ch4mseFhatx}
MSE(\widehat{F}(x))&=&E(\widehat{F}^2(x))-F^2(x)\nonumber\\
&=&\int _x^\infty \left[1-\frac{1}{A_n(t)}\sum_{y_0} \sum_{y_1}...\sum_{y_r} c(n-1,y_0,y_1,\ldots ,y_r)\right.\nonumber\\
&&\left.\times t^{\sum_{k=0}^r(k+1)y_k}\right.\nonumber\\
&&\left. \times \sum_{k=0}^r a_k t^{k}I_{x/t}\left((k+1),\sum_{k=0}^r (k+1)y_k\right)\right]^2\nonumber\\
&&\times f(t)dt-F^2(x)
\end{eqnarray}
Similarly, using Theorem $\ref{ch4ft}$, $\eqref{ch4oppecdf}$ in $\eqref{ch4mseFhatx}$, we can get the value of the MSE of UMVUE of the CDF.

\section{Particular case}\label{ch4secparticular}
\ In this section, we have studied in detail of the length-biased Lindley and Sujatha distribution that are particular cases of the OPPE distribution. The estimators of the PDF and the CDF are explicitly written, and their MSEs are compared.
Another particular case of the OPPE distribution is the Lindley distribution. The estimation of the PDF and the CDF of the Lindley distribution has been studied in detail in Maiti and Mukherjee (2018).

\subsection{Length-biased Lindley distribution}
Substituting $r=2,~a_0=0,~a_1=1,~a_2=1$ in $\eqref{ch4oppepdf}$ and $\eqref{ch4oppecdf}$, we can have the PDF and the CDF of the length-biased Lindley distribution respectively.\\
The PDF is
\begin{equation}\label{lbpdf}
f(x)=\frac{\theta^3}{2+\theta}x(1+x)e^{-\theta x};x,\theta>0
\end{equation}
and the CDF is
\begin{equation}\label{lbcdf}
F(x)= 1-\left[1+\frac{\theta x}{2+\theta}(2+\theta x+ \theta)\right]e^{-\theta x};x,\theta>0.
\end{equation}
\subsubsection{MLE of the PDF and the CDF}
\ If we substitute $r=2,~a_0=0,~a_1=1,~a_2=1$ in $\eqref{ch4mle}$ we will get the given expression as
\[
\frac{2\theta+6}{\theta(\theta+2)}-\bar{X}=0.
\]
Solving numerically the above mentioned equation, we can get the MLE of the length-biased Lindley distribution. 
 
\subsubsection{UMVUE of the PDF and the CDF}

\ In this section, we obtain the UMVUE of the PDF and the CDF of the length-biased Lindley distribution. Also, we obtain the MSEs of these estimators.



\vspace{0.2in}
To derive the UMVUE of the PDF and the CDF, we substitute $r=2,~a_0=0,~a_1=1,~a_2=1$ in Theorem $\eqref{ch4umvueepdfcdf}$.
\begin{t1}\label{umvuefandFlb}
Let $T=t$ be given. Then the UMVUE of the PDF is 
\begin{eqnarray*}
\widehat{f}(x)&=&\frac{p(x)}{A_{n}(t)}\sum_{y_1}\sum_{y_2} c(n-1,y_1,y_2)\\
&&\times (t-x)^{2y_1+3y_2-1},~0~<~x~<~t,
\end{eqnarray*}

and the UMVUE of the CDF is 

\begin{eqnarray*}
\widehat{F}(x)&=&1-\frac{1}{A_n(t)}\sum_{y_1}\sum_{y_2} c(n-1,y_1,y_2)t^{(2y_1+3y_2)} \\
&&\times \left[t I_{x/t}(2,2y_1+3y_2)+t^2 I_{x/t}(3,2y_1+3y_2)\right],~0~<~x~<~t.
\end{eqnarray*}
\end{t1}



\vspace{0.2in}
The MSE of the UMVUE of the PDF is given by
\begin{eqnarray}\label{ch4msefhatxlb}
MSE(\widehat{f}(x))&=&E(\widehat{f}^2(x))-f^2(x)\nonumber\\
&=&\int _x^\infty \left[\frac{p(x)}{A_{n}(t)}\sum_{y_1}\sum_{y_2} c(n-1,y_1,y_2)\right.\nonumber\\
&&\left.\times (t-x)^{2y_1+3y_2-1}\right]^2\nonumber\\
&&\times f(t)dt-f^2(x)
\end{eqnarray}
Using Theorem $\ref{ch4ft}$, $\eqref{lbpdf}$ in $\eqref{ch4msefhatxlb}$, we can get the value of the MSE of UMVUE of the PDF.

And the MSE of the UMVUE of the CDF is given by
\begin{eqnarray}\label{ch4mseFhatxlb}
MSE(\widehat{F}(x))&=&E(\widehat{F}^2(x))-F^2(x)\nonumber\\
&=&\int _x^\infty \left[1-\frac{1}{A_n(t)}\sum_{y_1}\sum_{y_2} c(n-1,y_1,y_2)t^{(2y_1+3y_2)} \right.\nonumber\\
&&\left.\times \left[t I_{x/t}(2,2y_1+3y_2)+t^2 I_{x/t}(3,2y_1+3y_2)\right]\right]^2\nonumber\\
&&\times f(t)dt-F^2(x)
\end{eqnarray}
Similarly, using Theorem $\ref{ch4ft}$, $\eqref{lbcdf}$ in $\eqref{ch4mseFhatxlb}$, we can get the value of the MSE of UMVUE of the CDF.

\begin{figure}[H]
\centering
\includegraphics[height=2.5in,width=6in]{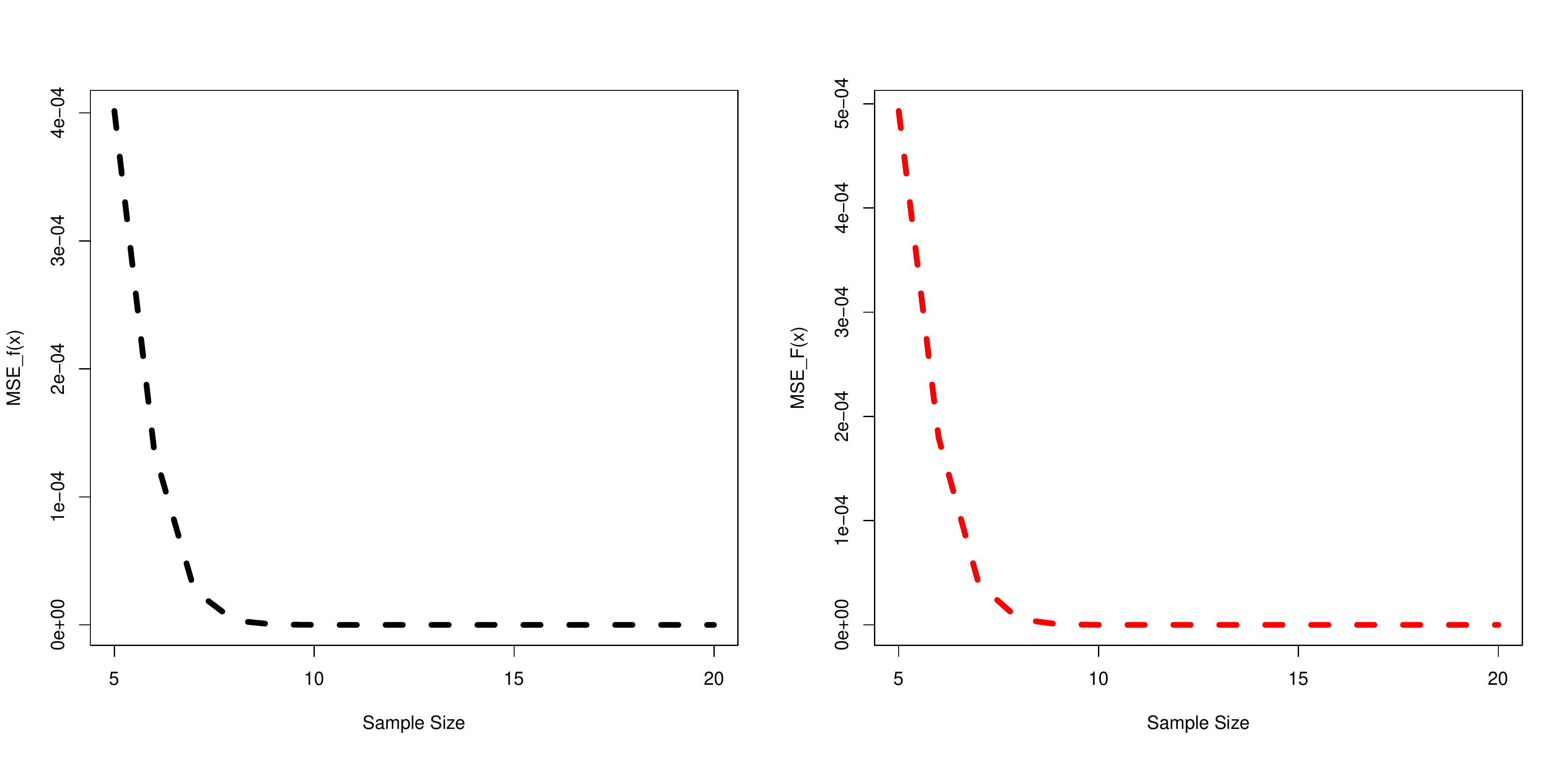}
\caption{Graph of theoretical MSE of UMVUE of the PDF and the CDF of the length-biased Lindley distribution for $\theta =0.1$, $x=2$ and $r=2$.}\label{ch4theorylb}
\end{figure}

Theoretical graph of the MSE of the UMVUE of the length-biased Lindley distribution is presented in Figure $\ref{ch4theorylb}$.

\subsection{Sujatha distribution}
\ By substituting $r=2,~a_0=1,~a_1=1,~a_2=1$ in $\eqref{ch4oppepdf}$ and $\eqref{ch4oppecdf}$, we will get the PDF and the CDF of the Sujatha Distribution respectively.\\
The PDF is
\begin{equation}\label{sujathapdf}
f(x)=\frac{\theta^3}{2+\theta+\theta ^2}(1+x+x^2)e^{-\theta x};x,\theta>0
\end{equation}
and the CDF is
\begin{equation}\label{sujathacdf}
F(x)=1-\left[1+\frac{\theta x}{2+\theta+\theta^2}(2+\theta x+ \theta)\right]e^{-\theta x};x,\theta>0.
\end{equation}
\subsubsection{MLE of the PDF and the CDF}
\ If we substitute $r=2,~a_0=1,~a_1=1,~a_2=1$ in $\eqref{ch4mle}$ we will get the given expression as
\[
\frac{\theta^2+2\theta+6}{\theta(\theta^2+\theta+2)}-\bar{X}=0.
\]
Solving numerically the above mentioned equation, we can get the MLE of the Sujatha distribution. 
 
\subsubsection{UMVUE of the PDF and the CDF}

\ In this section, we obtain the UMVUE of the PDF and the CDF of the Sujatha distribution. Also, we obtain the MSEs of these estimators.




\vspace{0.2in}
To derive the UMVUE of the PDF and the CDF, we use Theorem $\ref{ch4umvueepdfcdf}$ and replace $r=2,~a_0=1,~a_1=1,~a_2=1$.
\begin{t1}\label{umvuefandFsujatha}
Let $T=t$ be given. Then the UMVUE of the PDF is 
\begin{eqnarray*}
\widehat{f}(x)&=&\frac{p(x)}{A_{n}(t)}\sum_{y_0}\sum_{y_1}\sum_{y_2} c(n-1,y_0,y_1,y_2)\\
&&\times (t-x)^{y_0+2y_1+3y_2-1},~0~<~x~<~t,
\end{eqnarray*}

and the UMVUE of the CDF is 

\begin{eqnarray*}
\widehat{F}(x)&=&1-\frac{1}{A_n(t)}\sum_{y_0}\sum_{y_1}\sum_{y_2} c(n-1,y_0,y_1,y_2)t^{(y_0+2y_1+3y_2)} \\
&&\times \left[I_{x/t}(1,y_0+2y_1+3y_2)+t I_{x/t}(2,y_0+2y_1+3y_2)+t^2 I_{x/t}(3,y_0+2y_1+3y_2)\right],~0~<~x~<~t.
\end{eqnarray*}
\end{t1}



\vspace{0.2in}
The MSE of the UMVUE of the PDF is given by
\begin{eqnarray}\label{ch4msefhatxsujatha}
MSE(\widehat{f}(x))&=&E(\widehat{f}^2(x))-f^2(x)\nonumber\\
&=&\int _x^\infty \left[\frac{p(x)}{A_{n}(t)}\sum_{y_0}\sum_{y_1}\sum_{y_2} c(n-1,y_0,y_1,y_2)\right.\nonumber\\
&&\left.\times (t-x)^{y_0+2y_1+3y_2-1}\right]^2\nonumber\\
&&\times f(t)dt-f^2(x)
\end{eqnarray}
Using Theorem $\ref{ch4ft}$, $\eqref{sujathapdf}$ in $\eqref{ch4msefhatxsujatha}$, we can get the value of the MSE of UMVUE of the PDF.

And the MSE of the UMVUE of the CDF is given by
\begin{eqnarray}\label{ch4mseFhatxsujatha}
MSE(\widehat{F}(x))&=&E(\widehat{F}^2(x))-F^2(x)\nonumber\\
&=&\int _x^\infty \left[1-\frac{1}{A_n(t)}\sum_{y_0}\sum_{y_1}\sum_{y_2} c(n-1,y_0,y_1,y_2)t^{(y_0+2y_1+3y_2)} \right.\nonumber\\
&&\left.\times \left[I_{x/t}(1,y_0+2y_1+3y_2)+t I_{x/t}(2,y_0+2y_1+3y_2)+t^2 I_{x/t}(3,y_0+2y_1+3y_2)\right]\right]^2\nonumber\\
&&\times f(t)dt-F^2(x)
\end{eqnarray}
Similarly, using Theorem $\ref{ch4ft}$, $\eqref{sujathacdf}$ in $\eqref{ch4mseFhatxsujatha}$, we can get the value of the MSE of UMVUE of the CDF.

\begin{figure}[H]
\centering
\includegraphics[height=2.5in,width=6in]{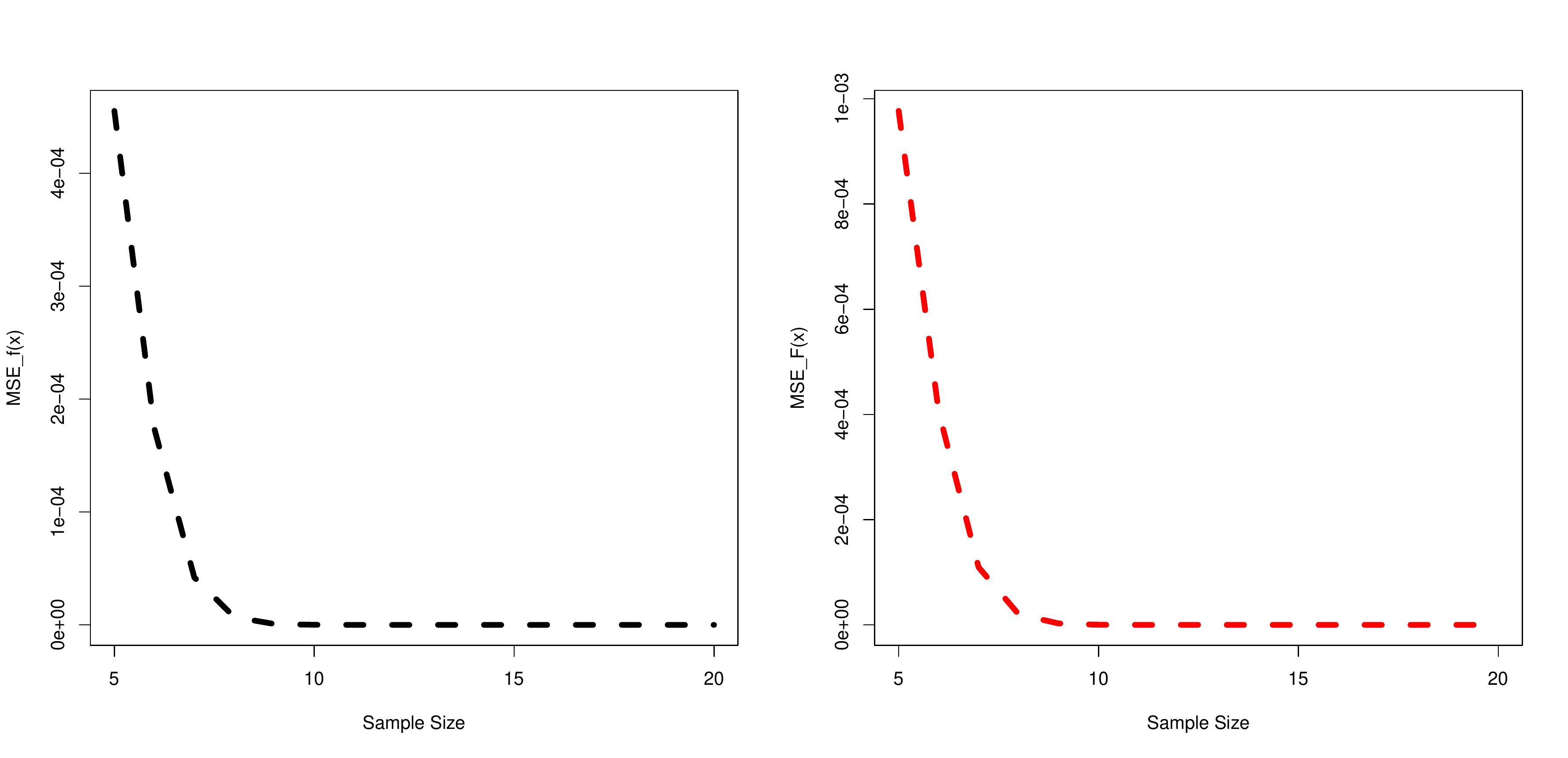}
\caption{Graph of theoretical MSE of UMVUE of the PDF and the CDF of the Sujatha distribution for $\theta =0.1$, $x=2$ and $r=2$.}\label{ch4theorysujatha}
\end{figure}

Theoretical graph of the MSE of the UMVUE of the Sujatha distribution is presented in Figure $\ref{ch4theorysujatha}$.

\section{Simulation}\label{ch4sim}
\ Direct application of Monte Carlo Simulation technique fails for generating random samples from the unified generalized Lindley, since the equation $$F(x)=u,~ u\in (0,1)$$ cannot be explicitly solved in $x$. On the other hand, one can use the fact that the distribution is a mixture of gamma distributions given in $\eqref{ch4oppepdf}$.\\
\par For the OPPE distribution the generation of random sample $X_1,~X_2,~\ldots,~X_n$ is distributed in the following algorithm:
\begin{enumerate}
\item[1.]Generate $U_{i}\sim Uniform (0,1),~ i=1(1)n$
\item[2.]If $\frac{\sum_{k=0}^{j-1}a_{k}\frac{k!}{\theta^{k+1}}}{\sum_{k=0}^{r}a_{k}\frac{k!}{\theta^{k+1}}}<U_{i}\leq \frac{\sum_{k=0}^{j}a_{k}\frac{k!}{\theta^{k+1}}}{\sum_{k=0}^{r}a_{k}\frac{k!}{\theta^{k+1}}},~j=1,2,...,r$,then set $X_{i}=V_{i}$,\\~where $V_{i}\sim gamma (j+1,\theta)$~and if $U_{i}\leq \frac{a_{0}~\frac{1}{\theta}}{\sum_{k=0}^{r}a_{k}\frac{k!}{\theta^{k+1}}}$, then set $X_{i}=V_{i}$,~where $V_{i}\sim\exp(\theta).$
\end{enumerate}

\begin{figure}[H]
\centering
\subfigure{\includegraphics[height=2.5in,width=2.9in]{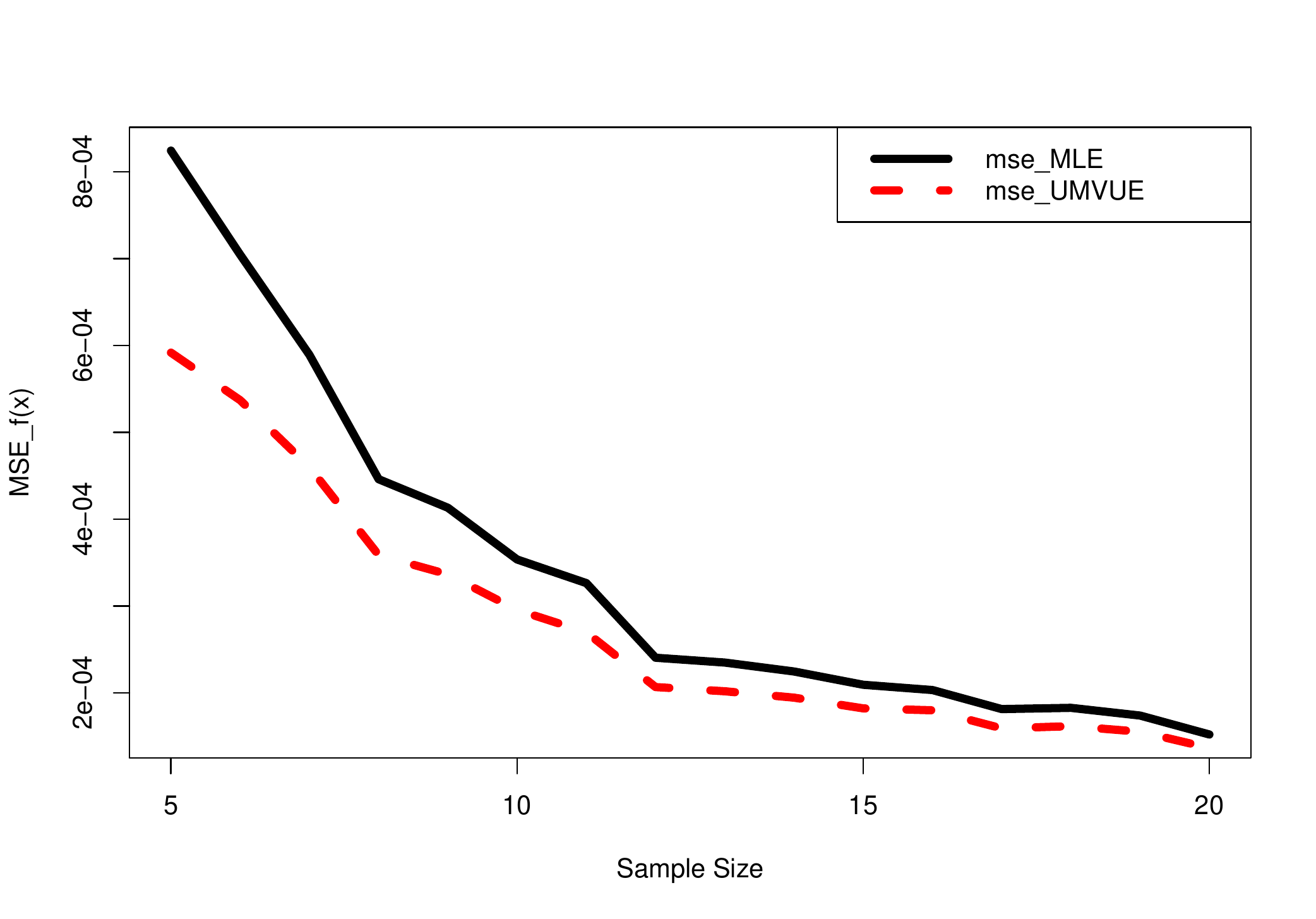}}
\subfigure{\includegraphics[height=2.5in,width=2.9in]{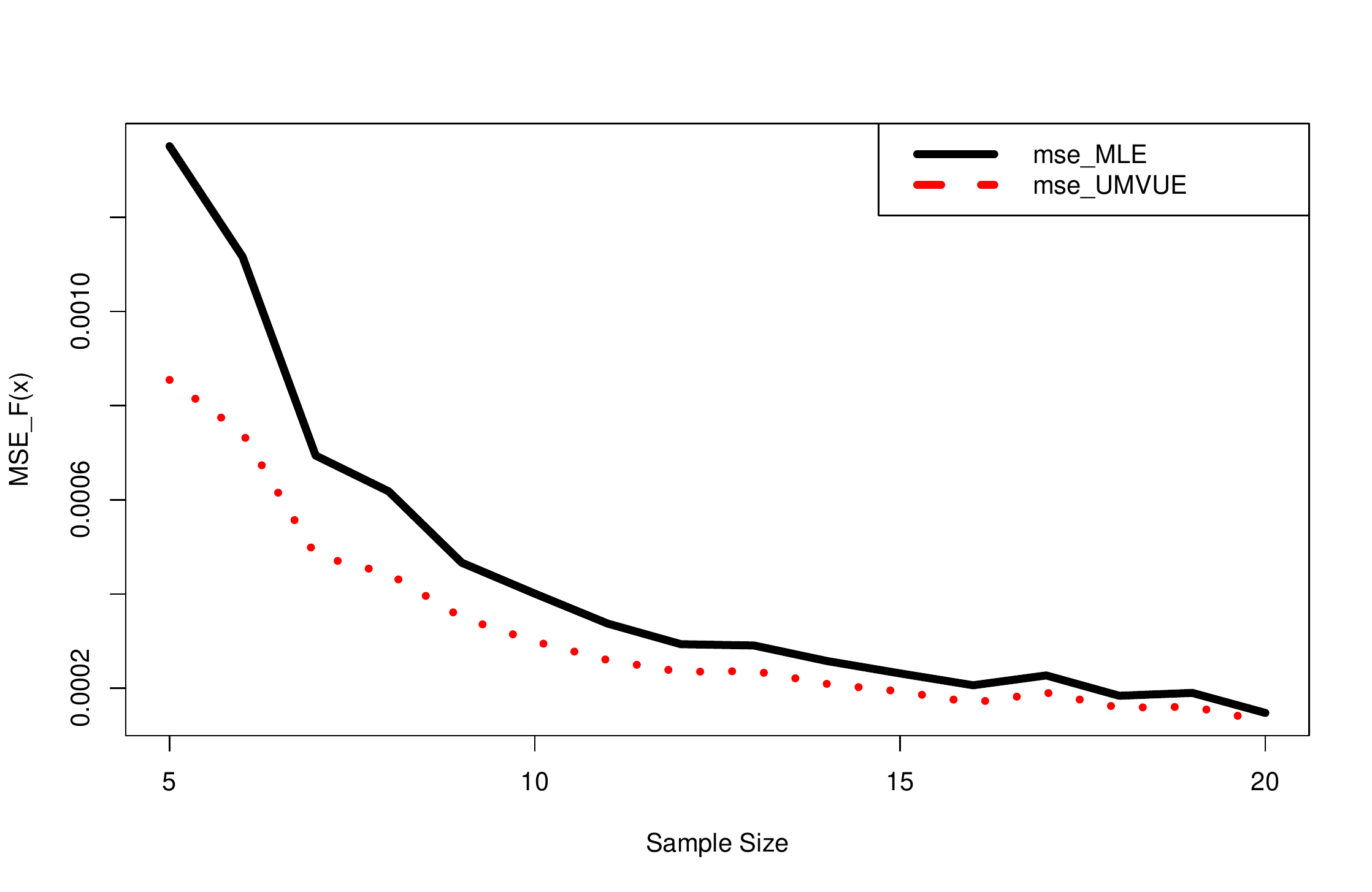}}
\caption{Graph of simulated MSE of the MLE of the PDF and the CDF of the length-biased Lindley distribution for $\theta =0.1$, $x=2$ and $r=2$.}\label{ch4simlb}
\end{figure}
\begin{figure}[H]
\centering
\subfigure{\includegraphics[height=2.5in,width=2.9in]{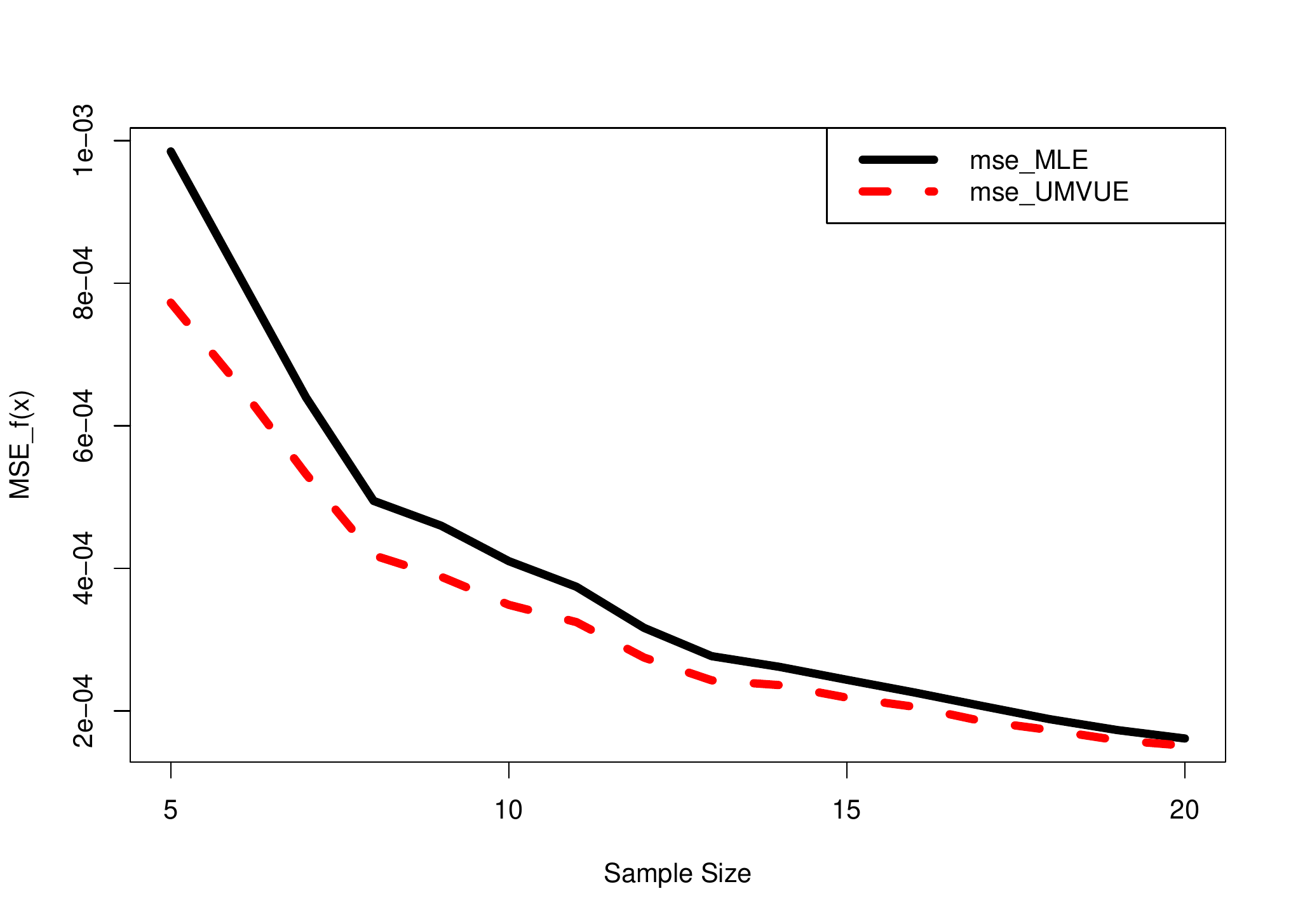}}
\subfigure{\includegraphics[height=2.5in,width=2.9in]{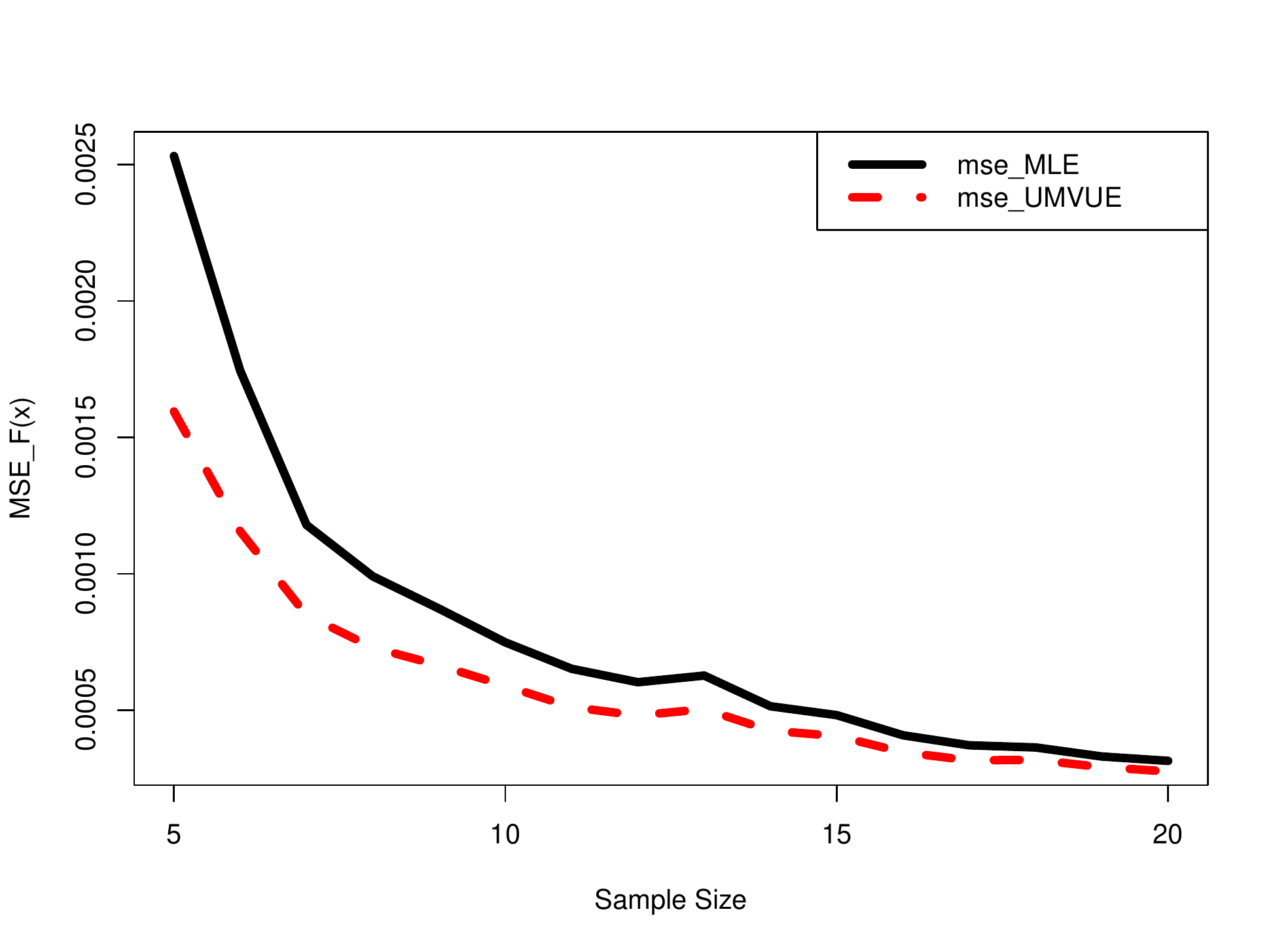}}
\caption{Graph of simulated MSE of the MLE of the PDF and the CDF of the Sujatha distribution for $\theta =0.1$, $x=2$ and $r=2$.}\label{ch4simsujatha}
\end{figure}

 A simulation is carried out with $1,000(N)$ repetitions. We choose $\theta = 0.1$, $x=2$ and $r = 2$ for both distribution. We compute MSE of the MLE and UMVUE of the PDF and the CDF. From Figures $\ref{ch4simlb}$ and $\ref{ch4simsujatha}$, it is clear that MSE decreases with increasing sample size that shows the consistency property of the estimators.

\section{Data Analysis}\label{ch4data}
\ In this section, we provide the analysis of two real data sets for comparing the performances of MLE and UMVUE for the PDF and the CDF. Table $\ref{ch4data3}$ represents the survival times (in days) of 72 guinea pigs infected with virulent tubercle bacilli, observed and reported by Elbatal et al.
(2013). Table $\ref{ch4data2}$ represents the failure times of the air conditioning system of an airplane and it is obtained from Linhart and Zucchini (1986).
\par
\begin{figure}[H]\label{ch4fit3}
\centering
\subfigure{\includegraphics[height=2.5in,width=2.9in]{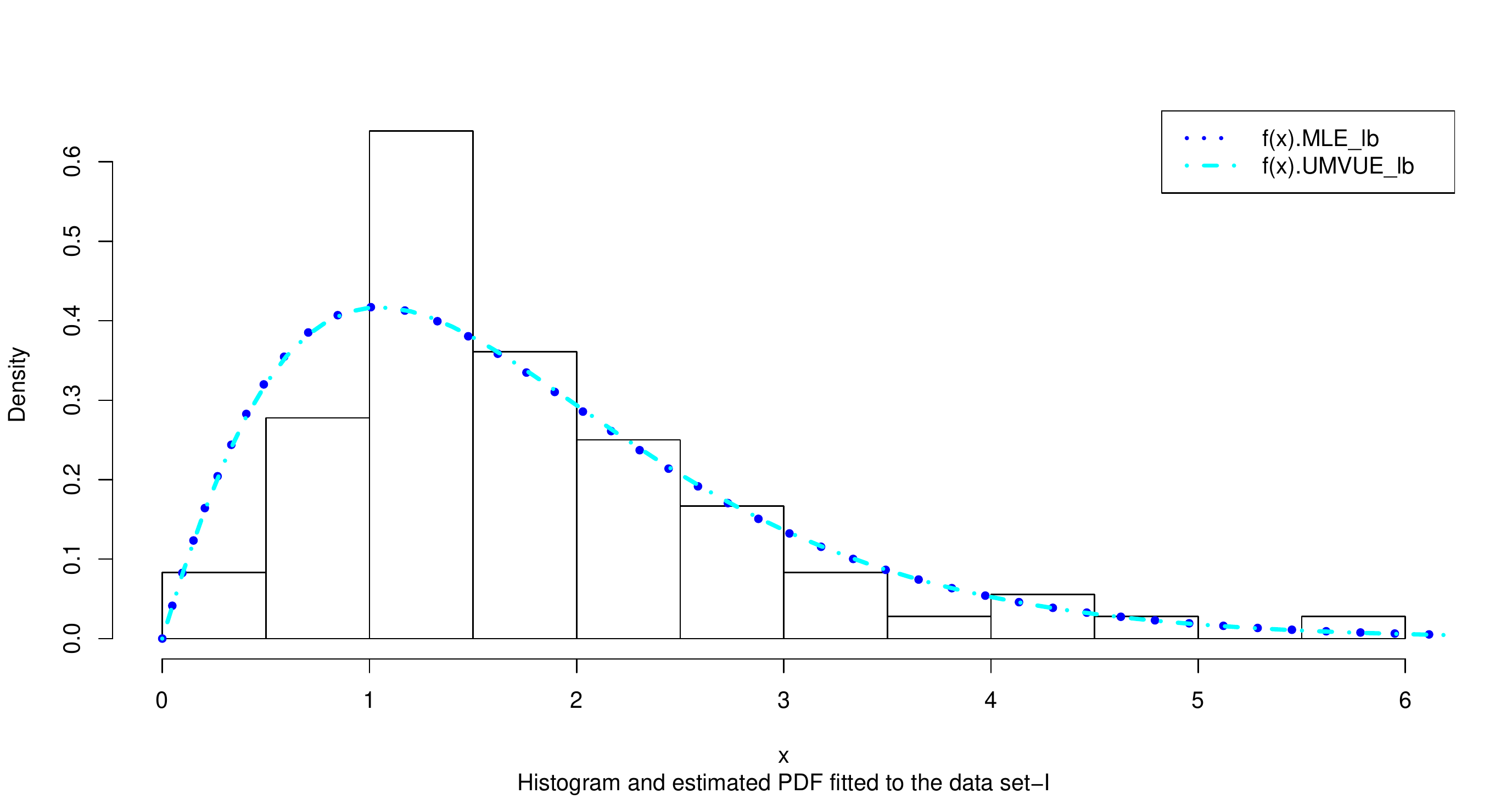}}
\subfigure{\includegraphics[height=2.5in,width=2.9in]{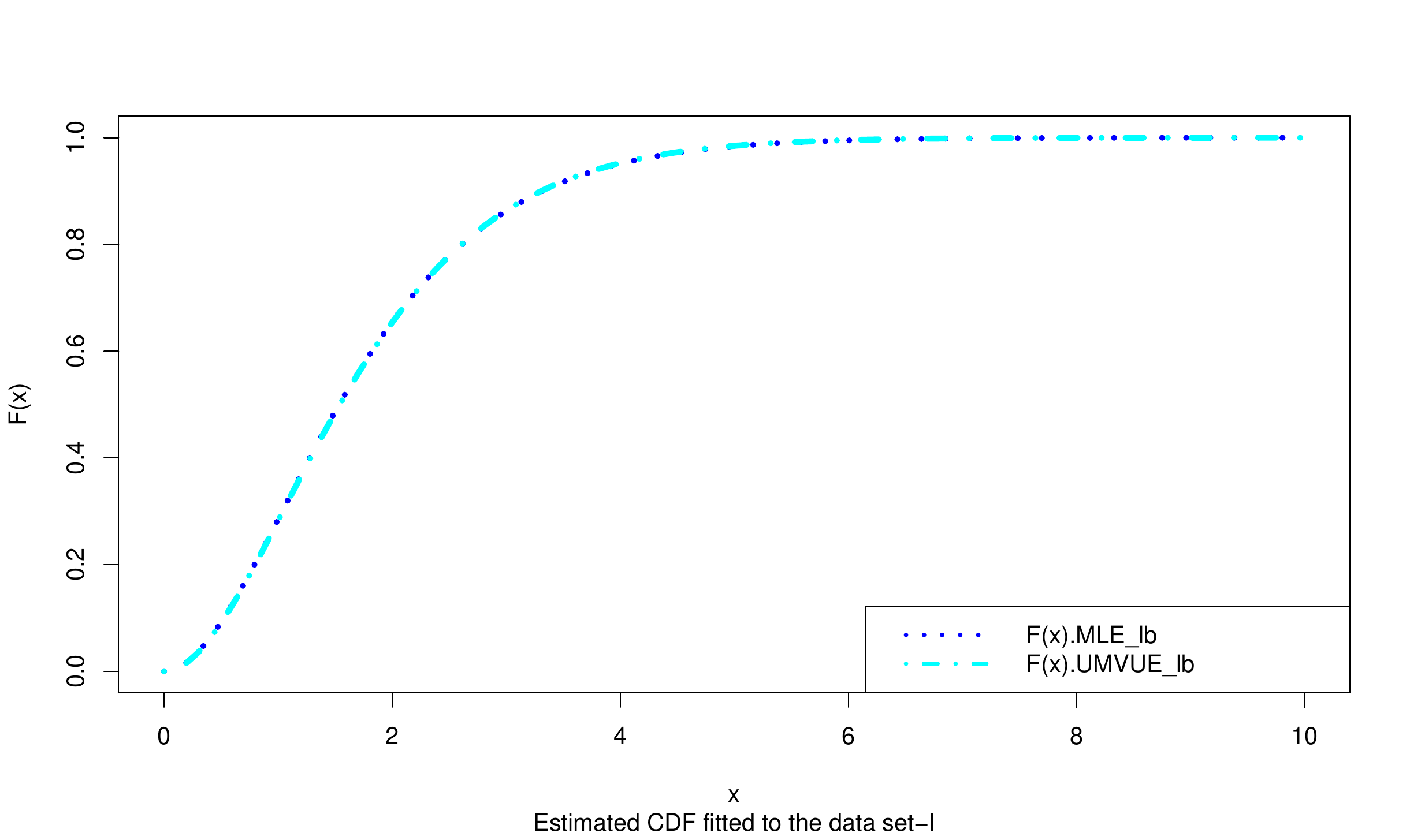}}
\caption{Graph of the estimated PDF and the CDF of length-biased Lindley distribution fitted to the data set-I.}
\end{figure}

 We fit length-biased Lindley distribution in data set-I (Table $\ref{ch4data3}$). The corresponding graph of histogram and the estimated PDF and the CDF has been shown in Figure $\ref{ch4fit3}$. From Figure $\ref{ch4fit3}$, we observe that the length-biased Lindley distribution shows good fit for data set-I. 
\par
\begin{figure}[H]\label{ch4fit2}
\centering
\subfigure{\includegraphics[height=2.5in,width=2.9in]{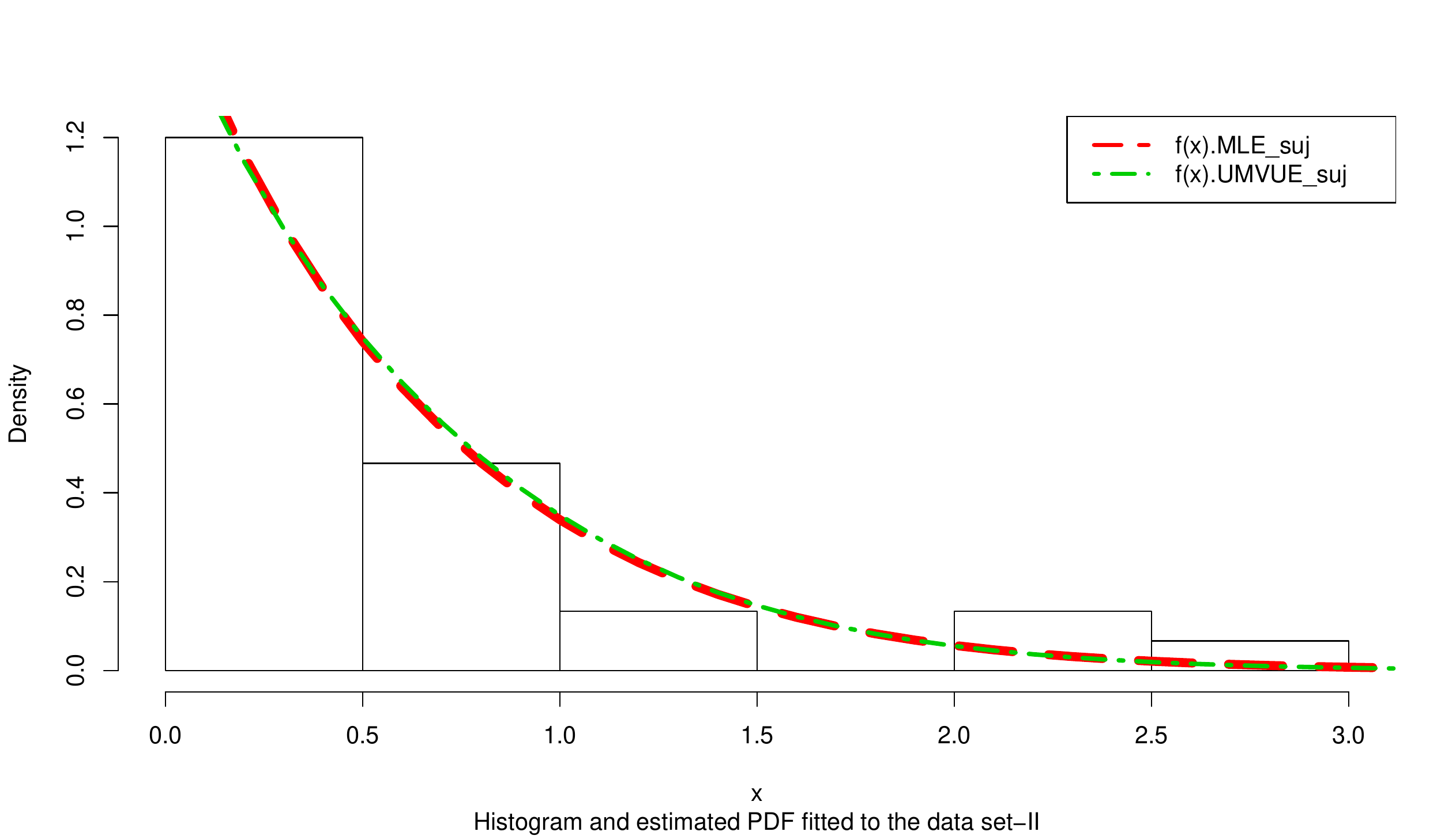}}
\subfigure{\includegraphics[height=2.5in,width=2.9in]{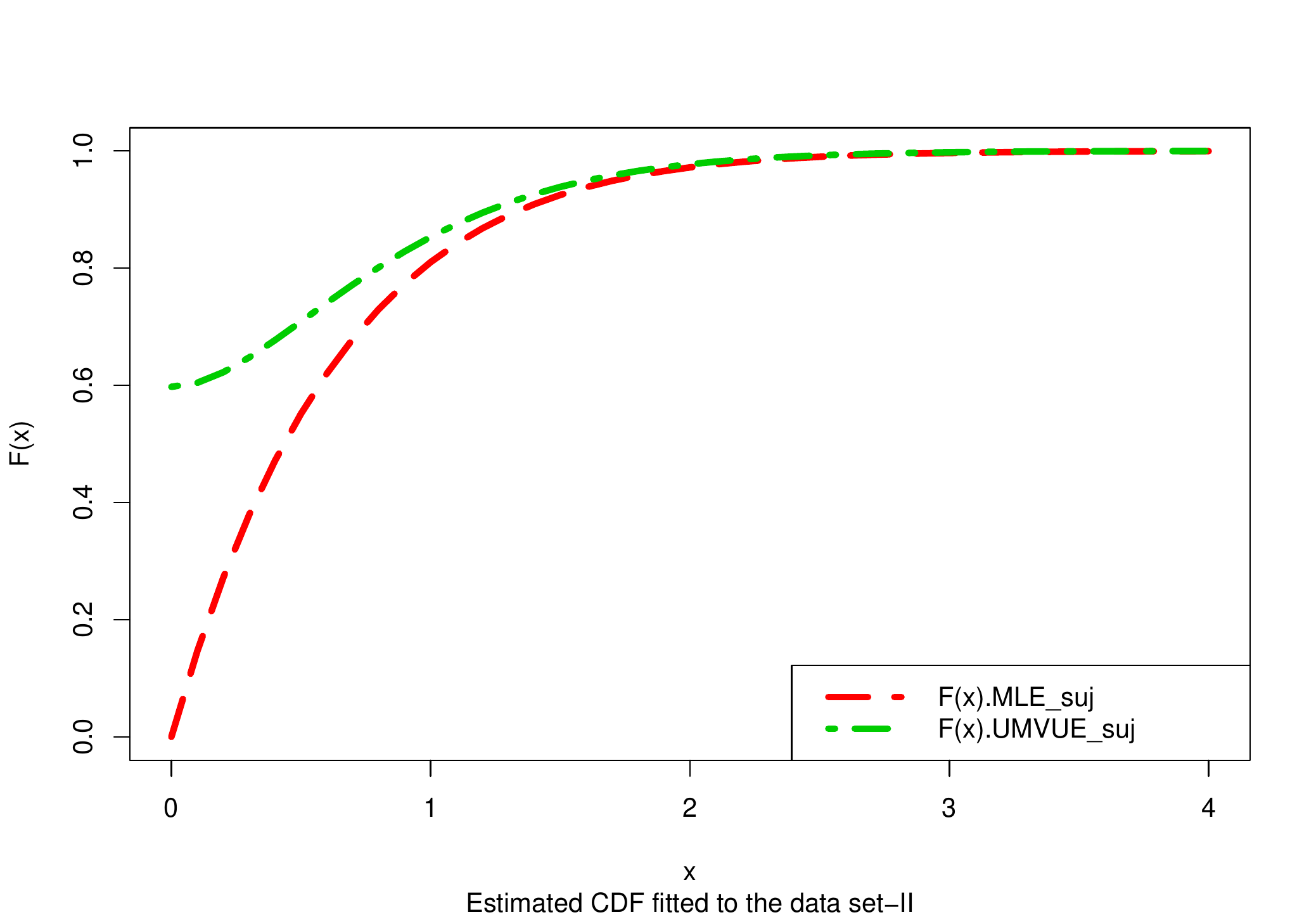}}
\caption{Graph of the estimated PDF and the CDF of Sujatha distribution fitted to the data set-II.}
\end{figure}

 Sujatha distribution have been fitted to data set-II (Table $\ref{ch4data2}$). Here, for computational ease, we have divided the whole data set by $100$. The graph of histogram and the estimated PDF and the CDF has been shown in Figure $\ref{ch4fit2}$. From Figure $\ref{ch4fit2}$, we observe that the Sujatha distribution shows good fit for data set-II.

\begin{table}[H]
\caption{\label{ch4data3} Survival times (in days) of 72 guinea pigs }
\begin{center}
{
\footnotesize 
\begin{tabular}{c c c c c c c c c}
\hline
0.1& 0.33& 0.44& 0.56& 0.59& 0.72& 0.74& 0.77& 0.92\\
 0.93& 0.96& 1& 1& 1.02& 1.05& 1.07& 1.07& 1.08\\
 1.08& 1.08& 1.09& 1.12& 1.13& 1.15& 1.16& 1.2& 1.21\\
 1.22& 1.22& 1.24& 1.3& 1.34& 1.36& 1.39& 1.44& 1.46\\
 1.53& 1.59& 1.6& 1.63& 1.63& 1.68& 1.71& 1.72& 1.76\\
1.83& 1.95& 1.96& 1.97& 2.02& 2.13& 2.15& 2.16& 2.22\\
2.3& 2.31& 2.4& 2.45& 2.51& 2.53& 2.54& 2.54& 2.78\\ 
2.93& 3.27& 3.42& 3.47& 3.61& 4.02& 4.32& 4.58& 5.55\\\hline
\end{tabular}
}
\end{center}
\end{table}

\begin{table}[H]
\caption{\label{ch4data2} Failure times of the air conditioning system of an airplane }
\begin{center}
{
\footnotesize 
\begin{tabular}{c c c c c c c c c c}
\hline
 23& 261& 87& 7 &120& 14& 62& 47& 225& 71\\
  246& 21& 42& 20& 5& 12& 120& 11& 3& 14\\ 
  71& 11& 14& 11& 16& 90& 1& 16& 52& 95\\\hline
\end{tabular}
}
\end{center}
\end{table}

\begin{table}[H]
\caption{\label{ch4model3}Model selection criterion-I}
\begin{center}
{
\footnotesize 
\begin{tabular}{|c|c|c|}
\hline
&\multicolumn{2}{|c|}{Negative log-likelihood}\\\cline{2-3}
Estimators &  \multicolumn{2}{|c|}{Length-biased Lindley distribution}\\\hline
MLE & \multicolumn{2}{|c|}{95.81244} \\\hline
UMVUE & \multicolumn{2}{|c|}{95.7132} \\\hline
\end{tabular}
}
\end{center}
\end{table}

\begin{table}[H]
\caption{\label{ch4model2}Model selection criterion-II}
\begin{center}
{
\footnotesize 
\begin{tabular}{|c|c|c|}
\hline
&\multicolumn{2}{|c|}{Negative log-likelihood}\\\cline{2-3}
Estimators &  \multicolumn{2}{|c|}{Sujatha distribution}\\\hline
MLE & \multicolumn{2}{|c|}{15.10749} \\\hline
UMVUE & \multicolumn{2}{|c|}{15.44566} \\\hline
\end{tabular}
}
\end{center}
\end{table}
\par Table $\ref{ch4model3}$ and $\ref{ch4model2}$ give the estimate of the negative log-likelihood values. Lower the value of negative log-likelihood indicates the better fit. From table $\ref{ch4model3}$ and $\ref{ch4model2}$, we see that the UMVUE and MLE is better in a negative log-likelihood sense, respectively.

\section{Concluding Remarks}\label{ch4conclusion}
\ Two estimators - MLE and UMVUE, has been found out for the PDF and the CDF of the length-biased Lindley and Sujatha distribution. The estimators have been compared theoretically as well as through simulation study in MSE sense. UMVUE is better for the PDF and the CDF in MSE sense for the length-biased Lindley and Sujatha distribution.


\end{document}